\documentclass{elsart}

\usepackage{amsmath}

\usepackage{amsbsy}
\usepackage{amssymb}
\usepackage[dvips]{graphicx}

\newcommand{\R}{\mathbb{R}}

\newcommand{\C}{\mathbb{C}}

\newcommand{\E}{\mathbb{E}}
\newcommand{\F}{\mathbb{F}}
\newcommand{\barf}{\mathbb{\overline{F}}}
\newcommand{\p}{\partial}
\newcommand{\ve}{\varepsilon}
\newcommand{\vp}{\varphi}

\newcommand{\lb}{\linebreak}
\newcommand{\br}[1]{|#1 \rangle}

\newcommand{\cl}[1]{\left\lceil #1 \right\rceil}

\newcommand{\inn}[2]{\langle #1| #2 \rangle}

\begin{document}

\begin{frontmatter}
  
  \title{Complexity of multivariate Feynman-Kac path integration in
      randomized \\ and quantum settings\thanksref{1}}

\thanks[1]{This research was supported in part 
  by the Defense Advanced Research Agency (DARPA) and Air Force
  Research Laboratory under agreement F30602-01-2-0523.}

\author{Marek Kwas} 
\ead{marek@cs.columbia.edu}

\address{
Department of Computer Science, Columbia University, \\ New York,
NY10027, USA,}
\address{
Institute of Applied Mathematics and Mechanics, 
University of Warsaw, \\
ul. Banacha 2, 02-097 Warszawa, Poland.}

\begin{abstract}
  The Feynman-Kac path integration problem was studied in the worst
  case setting by Plaskota et al.  (J. Comp. Phys.  164 (2000) 335)
  for the univariate case and by Kwas and Li (J. Comp. 19 (2003) 730)
  for the multivariate case with $d$ space variables.  In this paper
  we consider the multivariate Feynman-Kac path integration problem in
  the randomized and quantum settings. For smooth multivariate
  functions, it was proven in Kwas and Li (2003) that the classical
  worst case complexity suffers from the curse of dimensionality in
  $d$. We show that in both the randomized and quantum settings the
  curse of dimensionality is vanquished, i.e., the number of function
  evaluations and/or quantum queries required to compute an
  $\ve$-approximation has a bound independent of $d$ and depending
  polynomially on $\ve^{-1}$.  The exponents of these polynomials are
  at most 2 in the randomized setting and at most 1 in the quantum
  setting.  Hence we have exponential speedup over the classical worst
  case setting and quadratic speedup of the quantum setting over the
  randomized setting. However, both the randomized and quantum
  algorithms presented here still require extensive precomputing,
  similar to the algorithms of Plaskota et al. (2000) and Kwas and Li
  (2003).
\end{abstract}

\end{frontmatter}

\section{Introduction}
\label{sec:intro}
In this paper we study the multivariate Feynman-Kac path integration
problem.  Multivariate Feynman-Kac path integrals are path integrals
over the space of continuous functions from $\R_+$ to~$\R^d$, equipped
with a Wiener measure.  The multivariate Feynman-Kac path integral is
the solution of the initial value problem for the multivariate heat
equation, see Section \ref{sec:mvFK}. This type of path integral
plays a significant role in many fields, e.g., quantum physics and
chemistry, differential equations, and financial mathematics. A brief
survey of  research concerning path integrals is contained in the
introduction of \cite{PWW}.

In this paper we continue the research initiated in \cite{PWW}, where
a new algorithm for computing Feynman-Kac path integrals was proposed.
That paper dealt with the univariate case (i.e., with one space
variable) and the algorithm presented there was based on
$L_2$-approximation. An extension of the approach of \cite{PWW} to the
multivariate case (with many space variables) was presented in
\cite{KL}.  It turns out that algorithms based on $L_2$-approximation
are no longer applicable in the multivariate case.  The multivariate
case can be solved by using uniform approximation as a basic building
block.  In both papers \cite{KL,PWW} the problem was studied in the
worst case setting for input functions belonging to a class $\F$.
Assuming that the uniform approximation problem for the class $\F$ has
worst case complexity of order $\ve^{-\alpha(\F)}$ for some positive
$\alpha(\F)$ it was proved in \cite{KL} that the number of function
evaluations required to compute an $\ve$-approximation is roughly of
the same order $\ve^{-\alpha(\F)}$.  We stress that typically
$\alpha(\F)$ depends on the number $d$ of space variables and tends to
infinity with $d$, in which case we have the curse of dimensionality.

In this paper, we consider the multivariate Feynman-Kac path
integration problem in the   randomized and quantum settings.
We present algorithms that compute an $\ve$-approximation and analyze
their cost.  These algorithms are also based on uniform approximation.
However, the power of randomization and quantum computation permits
the improvement of the worst case complexity bound
$O(\ve^{-\alpha(\F)})$.  Namely, the number of function evaluations
required by the randomized algorithm is roughly of order $\ve^{-
  2\alpha(\F)/(\alpha(\F)+2)}$, whereas the number of function
evaluations and queries required by the quantum algorithm is roughly
of order $\ve^{- \alpha(\F)/(\alpha(\F)+1)}$, see Section
\ref{sec:error_analys}. We stress that the exponent of $\ve^{-1}$ in
the randomized setting is at most 2, and in the quantum setting is at
most 1.

In addition to providing the algorithms in the   randomized and
quantum settings we also study the complexity of multivariate
Feynman-Kac path integration in the randomized and quantum settings.
As in \cite{KL,PWW}, the complexity is bounded from below by the
complexity of multivariate weighted integration.  The upper bounds are
provided by the costs of the algorithms presented in this paper.

For the class $\F$ of $r$ times continuously differentiable functions
we have $\alpha(\F) = d/r$, and so the worst case setting suffers form
the curse of dimensionality. In the randomized setting, the complexity
is roughly of order $O(\ve^{-2/(1+2r/d)})$, whereas in the quantum
setting it is roughly of order $O(\ve^{-1/(1+r/d)})$. In
both cases the curse of dimensionality is vanquished. We thus have 
exponential speedup over the worst case setting. For $d \gg r$, we
have  quadratic speedup of quantum complexity over randomized
complexity.

\section{Multivariate Feynman-Kac path integration}
\label{sec:mvFK}

The multivariate Feynman-Kac formula is the solution of the initial
value problem for the heat (diffusion) equation
\begin{align}
  \label{eq:heat1}
  \frac{\p z}{\p t}(\mathbf{u},t)&=\tfrac{1}{2} \Delta z(\mathbf{u},t)
  + V(\mathbf{u})z(\mathbf{u},t)\qquad
  \mathrm{for}\; (\mathbf{u},t)\, \in \,\mathrm{int}\,\R^d \times
  [0,\infty),\\ 
  \label{eq:heat2}
  z(\mathbf{u},0)&=v(\mathbf{u}).
\end{align}
Here $v,V:\R^d\rightarrow\R$ are the initial value function and the
potential function, respectively. As usual, $\Delta$ denotes the
Laplacian. 
 
The solution $z$ of (\ref{eq:heat1}) and (\ref{eq:heat2}) is given by
the  Feynman-Kac formula
\begin{equation}
  \label{eq:fk-for}
  z(\mathbf{u},t) = \int_{\mathcal{C}} v(\mathbf{x}(t)+\mathbf{u}) \exp
  \left( \int_0^t V(\mathbf{x}(s)+\mathbf{u}) \, ds \right)
  w(d\mathbf{x}).
\end{equation}
Here, $\mathcal{C}$ is the set of continuous functions
$\mathbf{x}:\R_+\rightarrow\R^d$ such that $\mathbf{x}(0)=\mathbf{0}$.
The path integral~(\ref{eq:fk-for}) is with respect to the
$d$-dimensional Wiener measure~$w$, see \cite{Kar,RY}. Obviously,
(\ref{eq:fk-for}) holds for functions $v$ and $V$ for which the path
integral exists. In what follows, we assume that the functions $v$ and
$V$ belong to a class~$\F$ for which (\ref{eq:fk-for}) exists. This
class is  defined in Section \ref{sec:assum}.

\section{Problem}
\label{sec:problem}

For a given fixed point $(\mathbf{u}^*,t^*)\in \R^d \times [0,\infty)$
and arbitrary functions $v,V$ from the class $F$, we want to compute
an $\ve$-approximation of the exact solution
$z_{v,V}(\mathbf{u}^*,t^*) $ of~(\ref{eq:fk-for}).

The $\ve$-approximation $a_{v,V}(\mathbf{u}^*,t^*)$ is computed by an
algorithm $A_n$ that uses~$n$ function values of $v$ and $V$, i.e.,
$$
a_{v,V}(\mathbf{u}^*,t^*)=A_n\left( \mathbf{u}^*,t^*,
  v(\mathbf{u}_1), \ldots, v(\mathbf{u}_k),V(\mathbf{u}_{k+1}), \ldots,
  V(\mathbf{u}_n) \right).
$$

\subsection{Worst case setting}
\label{sec:wor-setting}

In the worst case setting the error of the algorithm $A_n$ is defined as
$$
e^\mathrm{wor}(A_n) = \sup_{v,V \in F}
|z_{v,V}(\mathbf{u}^*,t^*)-a_{v,V}(\mathbf{u}^*, t^*)|.
$$
We want  to determine the minimal number
$$
n^\mathrm{wor}(\ve, F)= \min \{ n: \; \exists\, A_n \quad \text{such
  that} \quad  e^\mathrm{wor}(A_n)
\le \ve \} 
$$
of function values that are needed to compute an
$\ve$-approximation in the worst case setting. This setting was
analyzed in \cite{KL}.

\subsection{Randomized setting}
\label{sec:rand}

In this setting we use randomized algorithms and replace the worst
case error assurance by an expected assurance. A randomized algorithm
$A_n$ depends on a random element $\omega$ chosen from some
probability space $\Omega$.  More precisely, we compute
\begin{equation}
\label{eq:gen-rand-algo}
a_{v,V}(\mathbf{u}^*,t^*; \omega)=A_{n,\omega}\left( \mathbf{u}^*,t^*,
  v(\mathbf{u}_{\omega,1}), \ldots,
  v(\mathbf{u}_{\omega,k}),V(\mathbf{u}_{\omega,k+1}), \ldots,
  V(\mathbf{u}_{\omega,n_\omega}) \right),
\end{equation}
with $n = \E_\omega(n_\omega)$. This means that we allow a random
choice of a mapping~$A_{n,\omega}$ and sample points $u_{\omega,i}$,
as well as the number $n_\omega$ of sample points, whose expected
value is fixed and equal to $n$.

We measure the randomized error of the algorithm $A_n$ with respect to
the $L_2$ norm, i.e.,
$$
e^\mathrm{rand}(A_n) := \sup_{v,V \in F}
\left(\E_\omega(z_{v,V}(\mathbf{u}^*,t^*) - 
  a_{v,V}(\mathbf{u}^*, t^*; \omega))^2 \right)^{1/2}.
$$
As before, we want to determine the  minimal expected
number of function  values 
$$
n^\text{rand}(\ve, F)=  \min \{ n: \; \exists \, A_n \quad \text{such
  that} \quad e^\mathrm{rand}(A_n)
\le \ve \} 
$$
needed to compute an $\ve$-approximation in the randomized setting.

\subsection{Quantum setting}
\label{sec:quant}

In the quantum setting we use quantum algorithms with (deterministic
or randomized) quantum queries and assume that we can also perform
function evaluations and arithmetic operations on a classical
computer.  These classical operations are used to prepare an input for
a quantum algorithm and to transform the outcome of a quantum algorithm
to an approximation of the exact solution.  We will be
interested in minimizing the total number of quantum queries and
function evaluations needed to compute an $\ve$-approximation.

In this section, we give a brief overview of a simplified quantum model
of computation for continuous problems and describe deterministic and
randomized quantum queries.  We refer the reader to
\cite{BHMT,H1,Stefannew,KW,NC,RandQQ} for  more detailed information.

We first outline a general framework of the quantum setting.
Assume that for a given class $H$ of input functions  $f: D
\rightarrow C$ we want to approximate the solution operator 
\begin{equation*}
  S: H \rightarrow G,
\end{equation*}
with $G$ being a normed space whose norm is denoted by $\|\cdot\|_G$.
We will approximate $S(f)$ by a quantum algorithm defined below.

First,  we transform a given input function $f \in
H$ by using a classical algorithm~$P_s$ with $s$ classical function
evaluations and obtain
$$
\bar f = P_s(f):D\rightarrow C.
$$
Then we use the transformed function $\bar f$ as an input to a
quantum algorithm.

Quantum algorithms are defined as follows. Let $\C^2$ be the two
dimensional complex vector space. Let $\mathcal{H}_k = \C^2 \otimes
\cdots \otimes \C^2$ be the $k$-fold tensor product of $\C^2$, having
dimension $2^k$.  Let $U_n(\bar f): \mathcal{H}_n \rightarrow
\mathcal{H}_n$ be a unitary operator of the form
\begin{equation*}
U_n(\bar f)= Q_n\, Q_{\bar f} Q_{n-1} \cdots Q_1\, Q_{\bar f} Q_0,
\end{equation*}
with unitary operators $Q_0,\ldots, Q_n$ and a quantum query
$Q_{\bar f}$, for some $\bar f \in P_s(H)$.

The deterministic quantum query $Q_{\bar f}$ is defined  as in
\cite{H1}.  Let $\mathcal{H}_k = \mathcal{H}_m \otimes
\mathcal{H}_{k-m}$ for some $m \le k$. Then $Q_{\bar f}: \mathcal{H}_m
\otimes \mathcal{H}_{k-m} \rightarrow \mathcal{H}_m \otimes
\mathcal{H}_{k-m}$ is a unitary operator of the form
\begin{equation}
\label{eq:det-qquery}
  Q_{\bar f} \br{x} \br{y} = \br{x} \big| y \oplus \beta ( \bar f
  ( \tau(x))) \big\rangle, 
\end{equation}
with
\begin{equation*}
  \tau:\{0,\ldots, 2^{k} -1\}\rightarrow D,\qquad \beta: C
  \rightarrow \{0, \ldots
  , 2^{k-m}-1\}
\end{equation*}
and $\oplus$ denoting the addition modulo $2^{k-m}$, see again
\cite{H1} for a more detailed discussion.

The randomized quantum query is defined in \cite{RandQQ}. In this
case, $Q_{\bar f} = Q_{\bar f,\,\omega}$ depends on a random element
$\omega$ and $Q_{\bar f,\,\omega}$ has the form (\ref{eq:det-qquery})
with $\tau=\tau_\omega$ depending on $\omega$. This permits the
computation of approximate values of $\bar f$ at randomized points.
Hence in this case the unitary operator $U_n$ depends on a random
element $\omega$ and has the form
\begin{equation*}
U_{n, \omega}( \bar f)= Q_{n_\omega}\, Q_{\bar
  f,\,\omega}\,Q_{n_\omega-1} \cdots Q_1 
Q_{\bar f,\,\omega}\, Q_0 \quad \text{with}\quad n=En_\omega.   
\end{equation*}
As usual we assume that the initial state is $\br{0}$ and we compute 
\begin{equation*}
  \br{\psi_{\bar f}} = U_{n}( \bar f) \br{0}=Q_n Q_{\bar
  f}Q_{n-1} \cdots Q_1 
  Q_{\bar f} Q_0 \br{0} 
\end{equation*}
for deterministic quantum queries and  
\begin{equation*}
  \br{\psi_{\bar f,\,\omega}}  = U_{n, \omega}( \bar f) \br{0}=Q_n
  Q_{\bar f,\,\omega}Q_{n-1} \cdots Q_1 
  Q_{\bar f,\,\omega} Q_0 \br{0}
\end{equation*}
for randomized quantum queries. Then we measure the final state and
obtain an outcome $j \in \{0,\ldots, 2^k -1\}$ with probability
\begin{equation*}
  p_{\bar f}(j) = |\inn{\psi_{\bar f}}{j}|^2 \quad
  \text{or}\quad  p_{\bar f,\,\omega}(j) = |\inn{\psi_{\bar
  f,\,\omega}}{j}|^2. 
\end{equation*}
Knowing the outcome $j$ we compute the final result on a classical
computer, and the quantum algorithm $A_n$ yields 
\begin{equation*}
    A_{n}(\bar f, j) = \phi(j)\qquad \text{or} \qquad
    A_{n, \omega}(\bar f, j) = \phi_\omega(j). 
\end{equation*}
for some $\phi$ or $\phi_\omega$.

In this paper we will be using quantum algorithms with randomized
quantum queries.  The error of such an algorithm $A_n$ is defined as
\begin{equation}
\label{eq:wor-avg-avg-error}
  e^\mathrm{quant}(A_n, P_s, S) = \sup_{f \in H} \left( \E_\omega 
  \E_\mathrm{q}\, \|
  S(f) - A_{n, \omega}(P_s (f), j)\|^2_G \right)^{1/2},
\end{equation}
where $\E_\omega$ is the expectation over the probability space
$\Omega$, and $\E_\mathrm{q}$ is the expectation with respect to
distribution of the quantum algorithm outcomes. 

Similarly to  the other settings, we want to determine the minimal
number of random quantum queries and classical function evaluations
\begin{equation*}
n^\text{quant}(\ve, H)= \min \{ s+n: \; \exists \, P_s
\;\exists\,A_n\quad \text{such that} \quad
e^\mathrm{quant}(A_n,P_s) \le \ve \}
\end{equation*}
needed to guarantee that the error does not exceed $\ve$. 
 
\begin{rem}
  We now briefly comment on the quantum error setting defined by
  (\ref{eq:wor-avg-avg-error}). Let us concentrate for a moment on the
  randomness introduced by a quantum algorithm, leaving aside
  randomized queries. So far, the literature dealing with continuous
  problems in the quantum setting has mainly considered probabilistic
  error.  That is, instead of taking an expectation with respect to
  all possible outcomes of a quantum algorithm (as $\E_\mathrm{q}$ in
  (\ref{eq:wor-avg-avg-error})), we want an error estimate such that
\begin{equation*}
 \left( \E_\omega  \|
  S(f) - A_{n, P_s(f), \omega}\|^2_G \right)^{1/2} \le \ve  
\end{equation*}
holds with a certain (high) probability, for any $f \in H$. Obviously
these two ways of measuring the error of a quantum algorithm are
related. We choose to study the average error for simplicity.
Moreover, the average error is probably more natural when we consider
randomized queries.
\end{rem}

The multivariate Feynman-Kac path integration problem
in the quantum setting is defined by taking $f=(v,V)$ with $H=F \times
F$ and $S(f)= z_{v,V}(\mathbf{u}^*,t^*)$.

\section{The function class $F$}
\label{sec:assum}

To assure the existence of the path integral
(\ref{eq:fk-for}), we need to choose a proper class of
input functions $F$, see \cite{KL}. We assume that 
\begin{equation}
  \label{eq:Fb}
  F=\{\,(f_1,f_2)\in \F\times \F : \quad \|f_1\|_\F\le \beta_1,\;\|f_2\|_\F\le
  \beta_2\,\}
\end{equation}
is a ball of a linear space $\F \times \F$ for some positive $\beta_1,
\beta_2$.

We make the following assumptions about the linear space $\F$. 
\begin{enumerate}
\item We assume that for every $\mathbf{u}\in \R^d$, the functional
  $L_{\mathbf{u}}:\F\rightarrow \R$ defined by
  $L_{\mathbf{u}}f=f(\mathbf{u})$ is continuous, and for arbitrary
  $a,t \in \R_+$ we have
\begin{equation}
\label{eq:cond}
\int_\mathcal{C} \|L_{\mathbf{x}(t)}\|_\F \exp \left( a \int_0^t
  \|L_{\mathbf{x}(s)}\|_\F\,ds\right)w(d\mathbf{x}) < \infty.
\end{equation}
By the Fernique theorem, see e.g.,~\cite{Kuo},
condition~(\ref{eq:cond}) holds if there exists $\alpha < 2$ such that
$\|L_{\mathbf{x}}\|_\F=O(\|\mathbf{x}\|^\alpha)$ for $\|\mathbf{x}\|$
approaching infinity, see \cite{PWW} for details.  Here and elsewhere,
$\|\mathbf{x}\|=\sqrt{\sum_{i=1}^d x_i^2}$ is the Euclidean norm
in~$\R^d$.

\item We assume that $\F$ is continuously embedded into
$L_\infty(\R^d)$.  That is, $\F \subset L_\infty(\R^d)$ and there
exists a positive $K$ such that
\begin{equation}
\label{eq:embed}
\|f\|_{L_\infty(\R^d)} \le K\|f\|_\F \qquad \forall f\in \F.
\end{equation}
This assumption permits us to relate the multivariate Feynman-Kac path
integration problem to uniform approximation  in the worst case
setting, see again \cite{KL}.  

By uniform approximation we mean the approximation of the embedding
operator $S:\F\rightarrow L_\infty(\R^d)$, $Sf=f$ in the norm of
$L_\infty(\R^d)$. Let $n^\text{wor}_\mathrm{APP}(\ve,\F)$ denote the
minimal number of function values needed to compute an
$\ve$-approximation in the worst case setting.  As we shall see in
Section \ref{sec:var}, uniform approximation also plays a significant
role in the randomized and quantum settings.

\item We assume that 
\begin{equation}
  \label{eq:unif_app_comp}
n^\text{wor}_\mathrm{APP}(\ve, \F)=O(\ve^{-\alpha(\F)})\qquad 
\mathrm{as}\quad  \ve \rightarrow 0,
\end{equation}
for some positive number $\alpha(\F)$.
\end{enumerate}

The linear space $\F$ is characterized by the exponent
$\alpha(\F)$. Usually $\alpha(\F)$ depends on the smoothness and the
number of variables of functions in $\F$, see Section \ref{sec:examples}.

\section{Feynman-Kac formula as a series of multivariate integrals}
\label{sec:series}

In this section we briefly recall  some results from \cite{KL} which
are needed for our analysis.

Without loss of generality we can assume $\mathbf{u}=\mathbf{0}$ in
(\ref{eq:fk-for}). Then we can express the path integral as a series
of multivariate integrals 
\begin{equation}
\label{eq:series}
S(v,V):= z(\mathbf{0},t) = \sum_{k=0}^{\infty}S_{k+1}(v,V),
\end{equation}
where
\begin{equation}
\label{eq:sk+1}
S_{k+1}(v,V)=\int_{\R^{(k+1)d}}v(\mathbf{z}_{k+1})\prod_{i=1}^k
V(\mathbf{z}_i)\, g_{k+1}(\mathbf{z}_1,\ldots,\mathbf{z}_{k+1})\,
d\mathbf{z}_1\ldots d\mathbf{z}_{k+1},
\end{equation}
with
\begin{equation}
\label{eq:g}
 g_{k+1}(\mathbf{z}_1,\ldots,\mathbf{z}_{k+1})
 =\int_{0\le t_1\le \cdots \le t_k \le t}
 f_{\,k+1}\,(t_1,\ldots,t_k,t,\mathbf{z}_1,\ldots,\mathbf{z}_{k+1})\,
 dt_1\ldots dt_k
\end{equation}
and
\begin{multline*}
  f_{\,k+1}\,(t_1,\ldots,t_k,t,\mathbf{z}_1,\ldots,\mathbf{z}_{k+1})
  =\left((2 \pi)^{k+1} t_1(t_2-t_1)\cdots(t-t_k)\right)^{-d/2}\\
  \times \exp\left(-\frac{1}{2}\left(\frac{\|\mathbf{z}_1\|^2}{t_1} +
      \frac{\|\mathbf{z}_2 -\mathbf{z}_1\|^2}{t_2-t_1}+ \cdots +
      \frac{\|\mathbf{z}_{k+1} -\mathbf{z}_k\|^2}{t-t_k}\right)
  \right).
\end{multline*}
Note that the integral (\ref{eq:sk+1}) depends on the input
functions $v$ and $V$ only through the product
\begin{equation*}
  h_{k+1}(\mathbf{z}_1,\ldots,\mathbf{z}_{k+1}) = v(\mathbf{z}_{k+1})
\prod_{i=1}^k V(\mathbf{z}_i)
\end{equation*}
and the weight functions $g_{k+1}$ can be computed in advance, albeit
with difficulty.  Let us recall also that
\begin{equation}
\label{eq:g_norm}
  \|g_{k+1}\|_{L_1(\R^{(k+1)d})}=\frac{t^k}{k!}\quad
\mathrm{for}\quad k \ge 0.
\end{equation}

\section{Approximation of one term of the series}
\label{sec:one}

In this section we present algorithms approximating one term of the
series (\ref{eq:series}). To make the notation more clear we define a
weighted integration operator
\begin{equation*}
  I_{k+1}(f) = \int_{\R^{(k+1)d}}
  f(\mathbf{z}_1,\ldots,\mathbf{z}_{k+1}) \;
  g_{k+1}(\mathbf{z}_1,\ldots,\mathbf{z}_{k+1})\, 
d\mathbf{z}_1\ldots d\mathbf{z}_{k+1}
\end{equation*}
where $f: \R^{(k+1)d} \rightarrow \R$ is an integrable function.
We can then rewrite one  term  of the series~(\ref{eq:series}) as
\begin{equation*}
S_{k+1}(v,V)=I_{k+1}(h_{k+1}).
\end{equation*}

In both the randomized and quantum settings, we shall use deterministic
uniform approximation of the function $h_{k+1}$.  To utilize the power
of randomization and/or quantum  computation, we will apply the
known technique of variance reduction.

\subsection{Variance reduction}
\label{sec:var}

Smolyak's algorithm is a powerful tool for computing an
$\ve$-approximation of tensor product problems.  For $h_{k+1} \in
\overbrace{\F \otimes\cdots \otimes \F}^{k+1}$, Smolyak's algorithm is
of the form
\begin{equation}
\label{eq:smolyak}
U_{\ve,k+1}(h_{k+1}) = \sum_{i=1}^{n(\ve,k+1)}
h_{k+1}(\mathbf{t}_{i,\ve,1}, 
\ldots, \mathbf{t}_{i,\ve,k+1}) 
\zeta_{i,\ve,k+1},  
\end{equation}
for some $\mathbf{t}_{i,\ve,j} \in \R^d$ and $\zeta_{i,\ve,k+1} \in
L_{\infty}(\R^{(k+1)d})$. It is proven in \cite[Lemma~2]{KL} that 
\begin{equation}
\label{eq:smolyak_error}
\|h_{k+1}-U_{\ve,k+1}(h_{k+1})\|_{L_{\infty}(\R^{(k+1)d})} \le \ve
\|v\|_\F\, \|V\|_\F^k, 
\end{equation}
where
\begin{equation}
\label{eq:tempcost}
  n(\ve,k+1) \le c_0 \bigg(c_1+c_2\frac{\ln
    1/\ve}{k}\bigg)^{(\alpha(\F)+1)k}_{+} 
\ve^{-\alpha(\F)},
\end{equation}
for some $c_i \in \R$. Here $a_{+}$ denotes $\max\{a,0\}$, the right
hand side of (\ref{eq:tempcost}) is defined to be $c_0 \,
\ve^{-\alpha(\F)}$  when
$k=0$.

The idea underlying variance reduction idea is as follows. First we
compute
$$
\bar h_{k+1, \ve} = U_{\ve,k+1}(h_{k+1})
$$
using $n(\ve, k+1)$ function values.
Then we compute
\begin{equation*}
  I_{k+1} (\bar h_{k+1, \ve}) = \sum_{i=1}^{n(\ve, k+1)}
  \bar h_{k+1}(\mathbf{t}_{i,\ve,1}, 
\ldots, \mathbf{t}_{i,\ve,k+1})
I_{k+1}( \zeta_{i,\ve,k+1}).  
\end{equation*}
Observe that the functions $\zeta_{i,\ve,k+1}$ do not depend on the
input functions $v$ and $V$ so the integrals
$I_{k+1}(\zeta_{i,\ve,k+1})$ can be precomputed.  

We stress that $\bar h_{k+1, \ve} $ and $ I_{k+1} (\bar h_{k+1, \ve})$
are deterministic.  We will use randomized or quantum algorithm to
approximate the multivariate integrals 
$$
I_{k+1}(h_{k+1} - \bar
h_{k+1, \ve}).
$$
Since the error depends on the norm $\|h_{k+1} - \bar h_{k+1,
  \ve}\|_{L_\infty (\R^{(k+1)d})}$, which is small, we can do this
efficiently. We present the details in the following two sections.

\subsection{Randomized algorithm}
\label{sec:ran_algo}

To make formulas simpler we define
\begin{equation*}
  \bar f_{k+1,\ve} = h_{k+1} - \bar h_{k+1, \ve}.
\end{equation*}
We use the randomized algorithm of the  form
\begin{equation}
  \label{eq:rand_algo}
  \phi^\mathrm{rand}_{\ve, m, \,k+1, \omega}(v,V)= I_{k+1}(\bar  h_{k+1, \ve})
+ Q^\mathrm{rand}_{m, k+1,\omega}(\bar f_{k+1,\ve}).
\end{equation}
Here 
\begin{equation}
  \label{eq:MC}
  Q^\mathrm{rand}_{m, k+1, \omega}(f) = \frac1{m} \sum_{j=1}^m f(x_{j,\omega})
\end{equation}
denotes the classical Monte Carlo algorithm with $m$ randomized sample
points.
 
Randomized  sample points are chosen with respect to the density \lb
$g_{k+1}/\|g_{k+1}\|_{L_1 (\R^{(k+1)d})}$ which is indicated by the
random parameter $\omega \in \Omega$.

Using the well known error formula for the classical Monte Carlo
algorithm, we conclude  that
\begin{multline}
  \label{eq:rand_err}
  \left( \E_\omega ( I_{k+1}( h_{k+1}) -
    \phi^\mathrm{rand}_{\ve,m,k+1, \omega}(v,V) )^2 \right)^{1/2} \\ 
  = \left( \E_\omega \left( I_{k+1} (\bar f_{k+1,\ve}) -
      Q^\mathrm{rand}_{m, k+1,\omega}(\bar f_{k+1,\ve}) \right)^2
  \right)^{1/2} \\ = \frac1{\sqrt{m}} \left(\mathrm{Var}(\bar
    f_{k+1,\ve})\right)^{1/2},
\end{multline}
with
\begin{equation*}
  \label{var}
  \mathrm{Var}( \bar f_{k+1,\ve}) =  I_{k+1}
   (\bar f_{k+1,\ve}^{\,2}) -
  \left(I_{k+1} (\bar f_{k+1,\ve})\right)^2.
\end{equation*}

Clearly, from (\ref{eq:smolyak_error}) and then from
(\ref{eq:Fb}), (\ref{eq:g_norm})  we get
\begin{equation}
  \left(\mathrm{Var}( \bar f_{k+1,\ve} ) \right)^{1/2}
  \le \frac{t^k}{k!}\; \|\bar f_{k+1,\ve}\|_{L_\infty (\R^{(k+1)d})} 
  \le \ve \; \frac{\|v\|_F\, \|V\|_F^k \, t^k}{k!} \le \ve \;
  \frac{\beta_1\, \beta_2^k\, t^k}{k!}.
\end{equation}
This yields the  error estimate
\begin{equation}
  \label{eq:rand-total-err}
  \left( \E_\omega ( I_{k+1}( h_{k+1}) - \phi^\mathrm{rand}_{\ve,\,k+1,
      \omega}(v,V) )^2 \right)^{1/2} \le   \frac{\ve}{\sqrt{m}} \;
      \frac{\beta_1\, \beta_2^k\, t^k}{k!} 
\end{equation}
and the total number of function evaluations is 
\begin{equation}
\label{eq:rand-cost-bound}
  n(\ve, k+1) + m.
\end{equation}

\subsection{Quantum algorithm}
\label{sec:quant_algo}
The structure of our quantum algorithm is similar to
randomized one, having the form
\begin{equation}
  \label{eq:quant_algo}
  \phi^\mathrm{quant}_{\ve,m,\kappa,\,k+1,  \omega}(v,V)=
  I_{k+1}( \bar  h_{k+1, \ve}) 
+ Q^\mathrm{quant}_{m,\kappa, k+1,d,\omega}(\bar f_{k+1,\ve}), 
\end{equation}
with, as before, $\bar f_{k+1,\ve} = h_{k+1} - \bar h_{k+1, \ve}$.
Here, we use a quantum algorithm $Q^\mathrm{quant}_{m, \kappa ,
  k+1}$, with $\kappa$ randomized quantum queries, that
approximates the classical Monte Carlo algorithm (\ref{eq:MC}). In
\cite{HKW} the problem of approximating 
\begin{equation*}
  \frac1{m} \sum_{j=1}^m f(x_{j,\omega})
\end{equation*}
was analyzed for Boolean functions $f$. Using the technique of
reducing the summation problem for bounded real functions to the
summation problem for Boolean functions as in \cite{H1}, we see that a
result similar to that of \cite{HKW} holds. From \cite{HKW} and
(\ref{eq:smolyak_error}) we conclude that
\begin{multline*}
   \left(  \E_\mathrm{q} \biggl(  \frac1{m} \sum_{j=1}^m f(x_{j,\omega}) -
  Q^\mathrm{quant}_{m,\kappa, k+1,\omega}(\bar f_{k+1,\ve})
  \biggr)^2 \right)^{1/2} \\ = O\left(\frac1{\kappa}  \|\bar
  f_{k+1,\ve}\|_{L_\infty (\R^{(k+1)d})}\right) = O\left(\frac{\ve}{\kappa}\,
  \beta_1 \beta_2^k\right).
\end{multline*}
By integrating  over $\Omega$, we obtain
\begin{multline}
\label{eq:quant-rand-err}
 \left( \E_\omega \E_\mathrm{q} \left|Q^\mathrm{rand}_{m,
    k+1,\omega}(\bar f_{k+1,\ve}) -
  Q^\mathrm{quant}_{m,\kappa, k+1,\omega}(\bar f_{k+1,\ve}) \right|^2
    \right)^{1/2} = O\left( \frac{\ve}{\kappa} \; \frac{\beta_1\, 
  \beta_2^k \, t^k}{k!}\right).
\end{multline}
 The total number of queries and
function evaluations is
\begin{equation*}
  n(\ve, k+1) + \kappa,
\end{equation*}
We stress that this number does not depend on $m$, which is only used
for the definition of the Monte Carlo algorithm. 

We now estimate the total error as 
\begin{multline*}
  \left(\E_\omega \E_\mathrm{q} ( I_{k+1}( h_{k+1}) -
    \phi^\mathrm{quant}_{\ve,m,\kappa, \,k+1,
      \omega}(v,V) )^2 \right)^{1/2} \\
  \le \left( \E_\omega \left( I_{k+1} (\bar f_{k+1,\ve}) -
      Q^\mathrm{rand}_{m, k+1,\omega}(\bar f_{k+1,\ve}) \right)^2
  \right)^{1/2} \\ + \left( \E_\omega \E_\mathrm{q}
    \left|Q^\mathrm{rand}_{m, k+1,\omega}(\bar f_{k+1,\ve}) -
      Q^\mathrm{quant}_{m,\kappa, k+1,\omega}(\bar f_{k+1,\ve})
    \right|^2 \right)^{1/2} .
\end{multline*}
This, by (\ref{eq:rand-total-err}) and (\ref{eq:quant-rand-err}), yields
\begin{equation*}
  \label{eq:quant-almost-total-err}
    \left(  \E_\omega \E_\mathrm{q}  ( I_{k+1}(h_{k+1}) -
    \phi^\mathrm{quant}_{\ve,m ,\kappa, \,k+1,
      \omega}(v,V) )^2 \right)^{1/2} = O\left(  \frac\ve{\sqrt{m}} \;
    \frac{\beta_1\, 
\beta_2^k\, t^k}{k!} + \frac\ve\kappa  \; \frac{\beta_1\,
    \beta_2^k \,t^k}{k!}\right).
\end{equation*} 
Letting $m = \kappa^2$ we get the error bound
\begin{equation}
  \label{eq:quant-total-err}
    \left(   \E_\omega \E_\mathrm{q} ( I_{k+1}( h_{k+1}) -
    \phi^\mathrm{quant}_{\ve,m,\kappa, \,k+1,
      \omega}(v,V) )^2 \right)^{1/2} = O\left( \frac\ve\kappa   \;
    \frac{2 \beta_1\, 
    \beta_2^k\, t^k}{k!} \right)
\end{equation} 
using
\begin{equation}
\label{eq:quant-cost-bound}
  n(\ve, k+1) + \kappa
\end{equation}
function values and quantum queries.  For the sake of convenience we denote
\begin{equation*}
  \phi^\mathrm{quant}_{\ve,\kappa, \,k+1,
      \omega}=\phi^\mathrm{quant}_{\ve,m,\kappa, \,k+1,
      \omega} \qquad \text{with $m=\kappa^2$}.
\end{equation*}

\section{Complete algorithms}
\label{sec:final_algo}

Based on the previous two sections we are ready to present algorithms
computing an $\ve$-approximation of multivariate Feynman-Kac path
integral. We  approximate consecutive terms of the series 
\begin{equation*}
  S(v,V)=\sum_{k=0}^{\infty}S_{k+1}(v,V)
\end{equation*}
by the algorithms
\begin{equation*}
  \phi^\mathrm{rand}_{\ve^{\mathrm{rand}}_{k+1},m_{k+1},k+1,
   \omega} \qquad \text{or}  \qquad
  \phi^\mathrm{quant}_{\ve^{\mathrm{quant}}_{k+1},\kappa_{k+1}, k+1,
  \omega},
\end{equation*}
with the accuracies $\ve^{\mathrm{rand}}_{k+1}$ and
$\ve^{\mathrm{quant}}_{k+1}$ in the corresponding settings being 
\begin{equation}
\label{eq:epsil}
  \ve^{\mathrm{rand}}_{k+1} = \ve^{2/(\alpha(\F)+2)} \frac{k!}{\beta_1
  \beta_2^k \, t^k 2^{k+1}}, \quad
  \ve^{\mathrm{quant}}_{k+1} = \ve^{1/(\alpha(\F)+1)} \frac{k!}{\beta_1 
  \beta_2^k \, t^k 2^{k+2}} 
\end{equation}
and the number of randomized sample points $m_{k+1}$ and quantum
queries $\kappa_{k+1}$ being
\begin{equation}
\label{eq:kappa&delta}
m_{k+1}= \cl{\ve^{-2\alpha(\F)/(\alpha(\F)+2)}} , \quad
  \kappa_{k+1} =  \cl{\ve^{- \alpha(\F)/(\alpha(\F)+1)}}.
\end{equation}

The final
forms of randomized and quantum algorithms approximating $S(v,V)$ are 
\begin{align*}
  \Phi^\mathrm{rand}_{\ve, \omega} (v,V) & = \sum_{k=0}^{N^\mathrm{rand}_\ve}
 \phi^\mathrm{rand}_{\ve^{\mathrm{rand}}_{k+1},m_{k+1}, \,k+1,
  \omega},\\ 
  \Phi^\mathrm{quant}_{\ve, \omega}(v,V) & = \sum_{k=0}^{N^\mathrm{quant}_\ve}
  \phi^\mathrm{quant}_{\ve^{\mathrm{quant}}_{k+1},\kappa_{k+1}, k+1,
  \omega},
\end{align*}
where the finite integers $N^\mathrm{rand}_\ve$ and
$N^\mathrm{quant}_\ve$ will be determined in the next section.

\subsection{Error analysis}
\label{sec:error_analys}

From (\ref{eq:rand-total-err}), (\ref{eq:quant-total-err}) and
(\ref{eq:epsil}), (\ref{eq:kappa&delta}), it is easy to check that
the  error bounds 
\begin{equation}
\label{eq:rand-sub-err}
  \left( \E_\omega ( I_{k+1}( h_{k+1}) -
  \phi^\mathrm{rand}_{\ve^{\mathrm{rand}}_{k+1},m_{k+1}, \,k+1, 
   \omega}(v,V) )^2 \right)^{1/2}  \le  \frac{\ve}{2^{k+1}}
\end{equation}
and
\begin{equation}
\label{eq:quant-sub-err}
    \left(   \E_\omega \E_\mathrm{q} ( I_{k+1}(h_{k+1}) -
    \phi^\mathrm{quant}_{\ve^{\mathrm{quant}}_{k+1},\kappa_{k+1}, k+1,
   \omega}(v,V) )^2 \right)^{1/2}  \le  \frac{\ve}{2^{k+1}}
\end{equation}
hold.

It is also easy to see that we need to approximate only a few terms.
Indeed, for~$k$ approaching infinity, we have
$\ve^{\mathrm{rand}}_{k+1}$ and $\ve^{\mathrm{quant}}_{k+1}$ also
tending to infinity.  Note that by (\ref{eq:embed}),
(\ref{eq:rand-total-err}) and (\ref{eq:quant-total-err}) we see that
for
\begin{equation*}
\frac{\ve^{\mathrm{rand}}_{k+1}}{m_{k+1}} \ge K^{k+1}\qquad \text{and}
\qquad \frac{\ve^{\mathrm{quant}}_{k+1}}{\kappa_{k+1}} \ge K^{k+1},  
\end{equation*}
 with $K$ being
the embedding constant in (\ref{eq:embed}),  the deterministic zero
algorithms provide sufficient accuracy.  Thus, we need to use the
algorithms $\phi^\mathrm{rand}_{\ve^{\mathrm{rand}}_{k+1},m_{k+1}, \,k+1,
   \omega}$ and $
 \phi^\mathrm{quant}_{\ve^{\mathrm{quant}}_{k+1},\kappa_{k+1}, k+1,
   \omega}$ 
only for $k=O(\ln(\ve^{-1}))$. Hence, we get $N^\mathrm{rand}_\ve =
O(\ln(\ve^{-1}))$ and $N^\mathrm{quant}_\ve = O(\ln(\ve^{-1}))$.

The bounds (\ref{eq:rand-sub-err}) and (\ref{eq:quant-sub-err}) yield
\begin{align}  
\label{eq:rand-tot-err}
  \left( \E_\omega ( S(v,V) - \Phi^\mathrm{rand}_{\ve,  \omega}(v,V) )^2
  \right)^{1/2} & \le \ve,\\ 
\label{eq:quant-tot-err}
    \left( \E_\omega \E_\mathrm{q} ( S(v,V) - 
    \Phi^\mathrm{quant}_{\ve,  \omega} (v,V) )^2 \right)^{1/2} & \le
    \ve. 
\end{align}
This means that the algorithms $\Phi^\mathrm{rand}_{\ve}$ and
$\Phi^\mathrm{quant}_{\ve}$ compute $\ve$-approximations of the
multivariate Feynman-Kac path integral in the randomized and quantum
settings respectively.

\subsection{Number of function values and quantum queries}
\label{sec:cost}
In this section we derive estimates on the numbers of function values
and quantum queries $n(\Phi^\mathrm{rand}_{\ve})$ and
$n(\Phi^\mathrm{quant}_{\ve})$ of the algorithms
$\Phi^\mathrm{rand}_{\ve}$ and $\Phi^\mathrm{quant}_{\ve}$. By the
bounds (\ref{eq:tempcost}), (\ref{eq:rand-cost-bound}) and
(\ref{eq:quant-cost-bound}) we get the obvious estimates
\begin{multline*}
  n\left(\Phi^\mathrm{rand}_{\ve}\right) = O\Bigg( \biggl(
    \beta_1^{\alpha(\F)} +  \sum_{k=1}^\infty
    \left(c_1+c_2\frac{\ln 1/\ve^\mathrm{rand}_{ k+1}
    }{k}\right)^{(\alpha(\F)+1)k}_{+} \\ \times
  \left(\frac{\beta_1  \beta_2^k t^k 2^{k+1}}{k!}\right)^{\alpha(\F)} +
    1 \biggr)  \ve^{- 2\alpha(\F)/(\alpha(\F)+2)}\Bigg) 
\end{multline*} 
and
\begin{multline*}
  n\left(\Phi^\mathrm{quant}_{\ve}\right) = O\Bigg(
    \biggl( \beta_1^{\alpha(\F)} + \sum_{k=1}^\infty
    \left(c_1+c_2\frac{\ln 1/\ve^\mathrm{quant}_{ k+1}
    }{k}\right)^{(\alpha(\F)+1)k}_{+}  \\ \times
  \left(\frac{\beta_1  \beta_2^k t^k 2^{k+1}}{k!}\right)^{\alpha(\F)}
     + 1 \biggr) \ve^{- \alpha(\F)/(\alpha(\F)+1) }\Bigg). 
\end{multline*} 
We can now use an argument similar to that in the proof of
\cite[Theorem 1]{PWW} to show that
\begin{align*}
  \sum_{k=1}^\infty
    \left(c_1+c_2\frac{\ln 1/\ve^\mathrm{rand}_{ k+1}
    }{k}\right)^{(\alpha(\F)+1)k}_{+} \left(\frac{\beta_1  \beta_2^k
    t^k 2^{k+1}}{k!}\right)^{\alpha(\F)} & = O(\ve^{-\delta}),\\
 \sum_{k=1}^\infty\left(c_1+c_2\frac{\ln 1/\ve^\mathrm{quant}_{k+1} 
    }{k}\right)^{(\alpha(\F)+1)k}_{+} \left(\frac{\beta_1  \beta_2^k
    t^k 2^{k+1}}{k!}\right)^{\alpha(\F)} & = O(\ve^{-\delta})
\end{align*} 
for all $\delta>0$. Thus  we finally get
\begin{align}
\label{eq:cost-phi-rand}
  n\left(\Phi^\mathrm{rand}_{\ve}\right) &= O\left(\ve^{-
      2\alpha(\F)/(\alpha(\F)+2) - \delta}\right),\\ 
\label{eq:cost-phi-quant}
n\left(\Phi^\mathrm{quant}_{\ve}\right) &= O\left(\ve^{-
      \alpha(\F)/(\alpha(\F)+1) - \delta}\right)
\end{align}
for all $\delta>0$.

\section{Complexity of multivariate Feynman-Kac path integration in
  randomized and quantum settings}
\label{sec:comp}

An analysis of the complexity of the multivariate Feynman-Kac path
integration in randomized and quantum settings is quite similar to the
one presented in \cite{KL} and \cite{PWW}. We only point out essential
differences.

\subsection{Lower bounds}
\label{sec:lower}

Lower bounds for our problem complexities are provided by the
complexities of multivariate weighted integration problem. By this
problem we mean an approximation of the integration operator
 $I:F \rightarrow \R $ define by 
\begin{equation*}
    I(f)=(2 \pi t^*)^{-d/2} \int_{\R^d} f(\mathbf{u})\,
    \exp(-\|\mathbf{u}\|/(2t^*)) \, d\mathbf{u} \qquad \forall\; f \in
    F. 
\end{equation*}
\vspace{12pt}
 
Consider a randomized algorithm $A^\mathrm{rand}_{n}$ that uses $n$
function values and approximates the integration operator $I$ . We say
that this algorithm computes an $\ve$-approximation of the weighted
integral if
\begin{equation*}
\label{eq:int-approx}
 \left( \E_\omega \left( I(f) -  A^\mathrm{rand}_{n,\omega} \right)^2
 \right)^{1/2} \le \ve  \qquad
 \forall\, f \in F. 
\end{equation*}
We denote by $n_\mathrm{INT}^\mathrm{rand}(\ve, \F)$ the minimal
number of function values needed to compute an $\ve$-approximation in
the randomized setting.

Consider a quantum algorithm $A^\mathrm{quant}_{n}$ that uses $n$
randomized quantum queries and approximates the operator $I$. We say
that $A^\mathrm{quant}_{n}$ computes an $\ve$-approximation of the
weighted integral if
\begin{equation}
\label{eq:int-quant-approx}
 \left( \E_\omega \E_\mathrm{q} \left( I(f) -  A^\mathrm{quant}_{n,
 \omega} \right)^2  \right)^{1/2} \le \ve  \qquad \forall \, f \in F.
\end{equation}
We define $n_\mathrm{INT}^\text{quant}(\ve, \F)$ as the minimal number
of quantum queries needed to compute an $\ve$-approximation.

As in \cite{KL},  we can reduce
multivariate Feynman-Kac path integration to multivariate integration
with a Gaussian weight by taking  $V \equiv 0$, since $ S(v,0) = I(v)$.
Moreover,    (\ref{eq:Fb})  and (\ref{eq:embed}) imply that 
\begin{align*}
  n_\mathrm{INT}^\text{rand}(\ve, \F) &\le n^\mathrm{rand}(\ve, \F),\\
  n_\mathrm{INT}^\text{quant}(\ve, \F) &\le n^\mathrm{quant}(\ve, \F).
\end{align*}

\subsection{Upper bounds}
\label{sec:upper}
Obvious estimates on the complexity of the multivariate Feynman-Kac
path integration are provided by the cost of the algorithms derived in 
Section \ref{sec:final_algo}. Thus, by (\ref{eq:cost-phi-rand}) and
(\ref{eq:cost-phi-quant})  we get
\begin{align*}
   n^\mathrm{rand}(\ve, \F) &= O\left(\ve^{-
      2\alpha(\F)/(\alpha(\F)+2) - \delta}\right),\\
 n^\mathrm{quant}(\ve, \F) &= O\left(\ve^{-
      \alpha(\F)/(\alpha(\F)+1) - \delta}\right)
\end{align*}
for all $\delta>0$, where $\alpha(\F)$ is the exponent of the uniform
approximation problem complexity for the space  $\F$ containing the
class $F$, i.e.,
\begin{equation*}
  n_\text{APP}^\text{wor} (\ve, \F) = O (\ve^{-\alpha(\F)}),
\end{equation*}
see  also Section \ref{sec:assum}.

From the previous two sections we can see when the randomized and quantum
algorithms proposed in this paper are almost optimal.  This is
the case for the classes of input functions for which randomized and
quantum complexities of the integration problem defined in Section
\ref{sec:lower} are of orders $\ve^{-2\alpha(\F)/(\alpha(\F)+2)}$ and
$\ve^{-\alpha(\F)/(\alpha(\F)+1)}$ respectively.

\section{Examples}
\label{sec:examples}

In this section we present two examples of function classes $F$
satisfying the assumptions from Section \ref{sec:assum} and compute
lower and upper bounds of of the complexities of the multivariate
Feynman-Kac path integration.

\subsubsection*{Weighted Sobolev space}

We use one of the results from \cite{WWwaord}, which relates the
complexity of the approximation of functions defined over a finite
domain to the complexity of the weighted approximation of functions
over the whole space $\R^d$. Let
\begin{equation*}
\barf = \left\{f:\R^d\rightarrow \R :\, f \in C^r(\R^d)\quad
\text{and}\quad \|f\|_{\barf}:=\sum_{0\le |\mathbf{a}|\le
  r}\|f^{(\mathbf{a})}\|_{L_\infty(\R^d)} < \infty \right\}.
\end{equation*}
For simplicity, we consider a weight function $\rho:\R^d\rightarrow
\R_+$, given by 
\begin{equation*}
\rho(\mathbf{z})=\exp(-\|\mathbf{z}\|^2) \qquad \forall z \in \R^d  
\end{equation*}
which decays exponentially.  By \cite{WWwaord} there exists an
algorithm 
\begin{equation*}
  U^\rho_\ve f=\sum_{i=1}^n f(\mathbf{t}_{i,\,\ve})\, \bar a_{i,\,\ve}
\end{equation*} 
that computes a weighted $\ve$-approximation of the function $f\in
\barf$, i.e.,
\begin{equation*}
  \|(f-U^\rho_\ve f)\,\rho\|_{L_\infty(\R^d)} \le \ve \,\|f\|_{\barf},
\end{equation*}
with 
\begin{equation*}
  n=O\left(\ve^{-d/r}\right).
\end{equation*}
Let 
\begin{equation*}
  \F=\{ f : \R^d\rightarrow \R:\; f/\rho \in \barf \quad \text{and}
  \quad \|f\|_\F:=\|f/ \rho\|_{\barf} < \infty \}.
\end{equation*}

We can use the algorithm $U^\rho_\ve$ to construct an algorithm
$U_\ve$ approximating functions from $\F$. Indeed, define
\begin{equation*}
  U_\ve f= \sum_{i=1}^n f(\mathbf{t}_{i,\,\ve})\, a_{i\,\ve},\quad
  \text{where} \quad
  a_{i,\,\ve}=\frac{\bar a_{i,\,\ve}\, \rho}{\rho(\mathbf{t}_{i,\,\ve})}. 
\end{equation*}
Then for $f_\rho=f/\rho$, we have
\begin{align*}
  \|f-U_\ve f\|_{L_\infty(\R^d)} = \|(f_\rho-U^\rho_\ve
  f_\rho)\,\rho\|_{L_\infty(\R^d)} \le \ve \, \|f_\rho\|_{\barf}=\ve
  \, \|f\|_\F,
\end{align*}
as claimed. We have to check the three remaining conditions which are
to be satisfied by $\F$, namely, the continuity of function evaluation
as well as conditions. It is
easy to see that for $f \in \F$ and $\mathbf{z}\in \R^d$ we have
\begin{equation*}
  f(\mathbf{z})\le \|f\|_{L_\infty(\R^d)} \le
\|f/\rho\|_{L_\infty(\R^d)} \le \|f\|_\F
\end{equation*}
and so function evaluation is continuous. Conditions (\ref{eq:cond})
and (\ref{eq:embed}) follow immediately from this continuity.

The algorithms $\Phi^\mathrm{rand}_{\ve}$ and
$\Phi^\mathrm{quant}_{\ve}$ compute an $\ve$-approximation of the
multivariate Feynman-Kac path integration problem for the class $F$
with the number of function evaluations and/or quantum queries roughly
$O(\ve^{-2/(1+2r/d)})$ and $O(\ve^{-1/(1+r/d)})$, respectively.  However,
the factors appearing in the big $O$ notation  depend on $d$ and
this dependence is exponential, see Sections \ref{sec:error_analys}
and \ref{sec:cost}. For $d \gg r$ the exponents $2d/(d+2r)$ for the
randomized algorithm and $d/(d+r)$ for the quantum algorithm are close
to $2$ and $1$.  In fact, the orders $2$ and $1$ can be obtained by
the use of the classical Monte Carlo algorithm (without variance
reduction). Then, the factors multiplying $\ve^{-2}$ and $\ve^{-1}$
are independent of $d$ for the class $F$ so the curse of
dimensionality present in the worst case setting (see \cite{KL}) is
indeed broken when we switch to the randomized or quantum settings.

To obtain lower bounds on the multivariate Feynman-Kac path
integration problem in the class $F$ we may switch to the integration
problem as in Section~\ref{sec:lower}. We observe that this
integration problem is not easier than the uniform integration over
the unit cube
\begin{equation*}
  I(f) = \int_{[0,1]^d} f(\mathbf{z})\, d\mathbf{z}
\end{equation*}
by taking functions with support $[0,1]^d$. It is known that the
uniform integration problem has the randomized complexity
$\Theta(\ve^{-2/(1+2r/d)})$, see \cite{stoch,Novak}, and quantum
complexity $\Theta(\ve^{-1/(1+r/d)})$, see \cite{RandQQ}. This shows
that the algorithms $\Phi^\mathrm{rand}_{\ve}$ and
$\Phi^\mathrm{quant}_{\ve}$ are roughly optimal for the class $F$.

\subsubsection*{Periodic functions}

This example was considered in \cite{KL}. We repeat all details for
the reader's convenience. Following \cite{Tem} we consider the class
$\barf$ of $2\pi$-periodic functions $f:[0,2\pi]^d\rightarrow \R^d$
satisfying the condition
\begin{multline}
  \label{eq:con}
  \forall \, f \in \barf \quad \, \forall j=1,\ldots,d \quad \exists\,
  \vp_j \in L_\infty([-2\pi,2 \pi]^d) \\ f(\mathbf{x})=\frac{1}{2 \pi}
  \int_0^{2 \pi} \vp_j(x_1,\ldots,x_j-t,\ldots,x_d)\,F_r(t)\,dt,
\end{multline}
where $r>0$ and
\begin{equation*}
  F_r(t)=1 + 2 \sum_{k=0}^\infty k^{-r}\, \cos\left(k\,t -
  \frac{r\,\pi}{2}\right).  
\end{equation*}
The norm in the class $\barf$ is defined as
\begin{equation*}
  \|f\|_{\barf}=\frac{1}{d}\,\sum_{j=1}^d
  \|\vp_j\|_{L_\infty([-2\pi,2 \pi]^d)}, 
\end{equation*}
where the $\vp_j$ are functions from the representation
(\ref{eq:con}) of the function $f$. In \cite{Tem}, there is  a
linear algorithm $U_\ve$ that computes a uniform $\ve$-approximation
of functions from the class~$\barf$, i.e., that 
\begin{equation*}
  \|f-U_\ve f\|_{L^\infty([0,2 \pi]^d)} \le \ve \, \|f\|_{\barf}
  \qquad \forall f \in \barf,
\end{equation*}
with the cost of order $\ve^{-d/r}$. 

Denote by $\F$ the class of functions $f:\R^d\rightarrow\R$ that are
periodic extensions of functions from $\barf$.  Let
$\|f\|_\F:=\|f|_{[0,2 \pi]^d}\|_{\barf}$.  Obviously, problem of the
\emph{uniform} approximation for the class~$\F$ can be obviously
solved using the algorithm mentioned above with the same cost as for
the class~$\barf$. Similarly to the previous example, we have to check
the three conditions of Section \ref{sec:assum}.  It is easy to see
that for $f \in \F$, $\mathbf{z} \in \R^d$, and arbitrary $j \in
\{1,2,\ldots,d\}$ we have
\begin{equation*}
  f(\mathbf{z}) \le \|f\|_{L^\infty(\R^d)}=
  \|f|_{[0,2\pi]^d}\|_{L^\infty([0,2\pi]^d)}\le C \, 
  \|\vp_j\|_{L^\infty([-2\pi,2\pi]^d)},
\end{equation*}
with $C=(2\pi)^{-1}\int_0^{2\pi}|F_r(t)|\, dt$. Hence
\begin{equation*}
  |f(\mathbf{z})|\le \|f\|_{L^\infty(\R^d)}\le C \, \|f\|_\F
\end{equation*}
and so function evaluation is continuous.  The remaining conditions
follow immediately.

Thus the algorithms $\Phi^\mathrm{rand}_{\ve}$ and
$\Phi^\mathrm{quant}_{\ve}$, based on the algorithm $U_\ve$ described
above, compute an $\ve$-approximation of the multivariate Feynman-Kac
path integral with a number of function evaluations and/or quantum
queries roughly $O(\ve^{-2/(1+2r/d)})$ and $O(\ve^{-1/(1+r/d)})$,
respectively.  Using an  argument similar to that of  the previous
example, we conclude that the algorithms $\Phi^\mathrm{rand}_{\ve}$ and
$\Phi^\mathrm{quant}_{\ve}$ are roughly optimal for the class $F$.

\begin{ack}
I wish to thank my advisor
H. Wo\'zniakowski for many inspiring discussions. I am also grateful
to S. Heinrich, J. F. Traub, A. G. Werschultz for valuable comments
and remarks.
\end{ack}

\end{document}